\documentclass[11pt]{article}
\usepackage{appb,epsfig}

\def\beq{\begin{equation}}
\def\eeq{\end{equation}}
\def\beqar{\begin{eqnarray}}
\def\eeqar{\end{eqnarray}}
\def\barr#1{\begin{array}{#1}}
\def\earr{\end{array}}
\def\bfi{\begin{figure}}
\def\efi{\end{figure}}
\def\btab{\begin{table}}
\def\etab{\end{table}}
\def\bce{\begin{center}}
\def\ece{\end{center}}

\def\text{\textstyle}

\def\al{\alpha}

\def\Ga{\Gamma}
\def\De{\Delta}

\def\refeq#1{\mbox{eq.~(\ref{#1})}}
\def\reffi#1{\mbox{Fig.~\ref{#1}}}

\def\refta#1{\mbox{Tab.~\ref{#1}}}
\def\citere#1{\mbox{Ref.~\cite{#1}}}

\newcommand{\GeV}{\unskip\,\mathrm{GeV}}
\newcommand{\MeV}{\unskip\,\mathrm{MeV}}
\newcommand{\TeV}{\unskip\,\mathrm{TeV}}

\def\mathswitchr#1{\relax\ifmmode{\mathrm{#1}}\else$\mathrm{#1}$\fi}

\newcommand{\PW}{\mathswitchr W}
\newcommand{\PZ}{\mathswitchr Z}

\newcommand{\PH}{\mathswitchr H}

\newcommand{\Pt}{\mathswitchr t}

\def\mathswitch#1{\relax\ifmmode#1\else$#1$\fi}

\newcommand{\MW}{\mathswitch {M_\PW}}

\newcommand{\MZ}{\mathswitch {M_\PZ}}
\newcommand{\MH}{\mathswitch {M_\PH}}

\newcommand{\Mt}{\mathswitch {m_\Pt}}
\newcommand{\scrs}{\scriptscriptstyle}
\newcommand{\sw}{\mathswitch {s_{\scrs\PW}}}

\newcommand{\cw}{\mathswitch {c_{\scrs\PW}}}
\newcommand{\sweff}{\sin^2 \theta_{\mathrm{eff}}}
\newcommand{\sweffsub}{\sin^2 \theta_{\mathrm{eff, subtr}}}

\newcommand{\MWsub}{M_{\PW, \mathrm{subtr}}}

\newcommand{\GF}{\mathswitch {G_\mu}}

\newcommand{\lsim}
{\;\raisebox{-.3em}{$\stackrel{\displaystyle <}{\sim}$}\;}

\newcommand{\fea}{{\em FeynArts}}

\newcommand{\Dri}[1]{\Delta{\mathswitchr r}_{\mathrm{#1}}}

\newcommand{\Da}{\Delta\alpha}
\newcommand{\Drho}{\Delta\rho}

\def\draftdate{\relax}
\def\mda{\relax}
\def\mua{\relax}
\def\mla{\relax}
\def\draft{
\def\thtystars{******************************}
\def\sixtystars{\thtystars\thtystars}
\typeout{}
\typeout{\sixtystars**}
\typeout{* Draft mode!
         For final version remove \protect\draft\space in source file
*}
\typeout{\sixtystars**}
\typeout{}
\def\draftdate{\today}
\def\mua{\marginpar[\boldmath\hfil$\uparrow$]%
                   {\boldmath$\uparrow$\hfil}%
                    \typeout{marginpar: $\uparrow$}\ignorespaces}
\def\mda{\marginpar[\boldmath\hfil$\downarrow$]%
                   {\boldmath$\downarrow$\hfil}%
                    \typeout{marginpar: $\downarrow$}\ignorespaces}
\def\mla{\marginpar[\boldmath\hfil$\rightarrow$]%
                   {\boldmath$\leftarrow $\hfil}%
                    \typeout{marginpar:
$\leftrightarrow$}\ignorespaces}
\def\Mua{\marginpar[\boldmath\hfil$\Uparrow$]%
                   {\boldmath$\Uparrow$\hfil}%
                    \typeout{marginpar: $\Uparrow$}\ignorespaces}
\def\Mda{\marginpar[\boldmath\hfil$\Downarrow$]%
                   {\boldmath$\Downarrow$\hfil}%
                    \typeout{marginpar: $\Downarrow$}\ignorespaces}
\def\Mla{\marginpar[\boldmath\hfil$\Rightarrow$]%
                   {\boldmath$\Leftarrow $\hfil}%
                    \typeout{marginpar:
$\Leftrightarrow$}\ignorespaces}
\overfullrule 5pt
\oddsidemargin -15mm
\marginparwidth 29mm
}

\begin{document}
\null
\hfill KA-TP-11-1998\\
\null
\hfill hep-ph/9807222\\
\vskip .8cm
\begin{center}
{\Large \bf Results for Precision Observables\\[.5em]
in the Electroweak Standard Model\\[.5em]
at Two-Loop Order and Beyond
}
\vskip 2.5em
{\large
{\sc Georg Weiglein}\\[1ex]
{\normalsize \it Institut f\"ur Theoretische Physik, Universit\"at
Karlsruhe,\\
D-76128 Karlsruhe, Germany}
}
\vskip 2em
\end{center} \par
\vskip 1.2cm
\vfil
{\bf Abstract} \par
Higher-order contributions to the precision observables $\De r$,
$\sweff$ and $\Ga_l$ are discussed. The Higgs-mass dependence of the 
observables is investigated at the two-loop level, and exact results are
derived for the Higgs-dependent two-loop corrections associated with fermions.
The top-quark corrections are compared with the results obtained by an
expansion in the top-quark mass up to
next-to-leading order. For the pure fermion-loop contributions to 
$\De r$ results up to four-loop order are derived. They allow to 
investigate the validity of the commonly used resummation of the 
leading fermionic contributions to $\De r$.
\par
\vskip 1cm
\null
\setcounter{page}{0}
\clearpage

\title{
RESULTS FOR PRECISION OBSERVABLES\\
IN THE ELECTROWEAK STANDARD MODEL\\
AT TWO-LOOP ORDER AND BEYOND%
\thanks{Presented at the Zeuthen
Workshop on Elementary Particle Physics, Loops and Legs in Gauge
Theories, Rheinsberg, April 19--24, 1998.
}
}
\author{Georg Weiglein
\address{Institut f\"ur Theoretische Physik, Universit\"at Karlsruhe,\\
D--76128 Karlsruhe, Germany}
}
\maketitle
\begin{abstract}
Higher-order contributions to the precision observables $\De r$,
$\sweff$ and $\Ga_l$ are discussed. The Higgs-mass dependence of the 
observables is investigated at the two-loop level, and exact results are
derived for 
the Higgs-dependent two-loop corrections associated with fermions.
The top-quark corrections are
compared with the results obtained by an
expansion in the top-quark mass up to
next-to-leading order. For the pure fermion-loop contributions to 
$\De r$ results
up to four-loop order are derived. They allow to investigate the
validity of the commonly used resummation of the leading fermionic
contributions to $\De r$.
\end{abstract}
  
\section{Introduction}

Confronting the predictions of the electroweak Standard Model (SM) with
the precision data~\cite{datamor98} allows to derive indirect constraints
on the mass of the Higgs boson, which is the last missing ingredient of
the minimal SM, and also sets the basis for the investigation of possible
effects of new physics via radiative corrections. The bounds on the
Higgs-boson mass, $\MH$, obtained from fitting the SM predictions to the
data, are strongly affected by the theoretical uncertainty due to
unknown higher-order corrections. In particular the two-loop electroweak
corrections have to be under control in order to obtain reliable bounds.
At this order, the resummations of the leading one-loop contributions
are known~\cite{resum}, the leading and next-to-leading term in an
expansion for asymptotically large values of the top-quark mass, $\Mt$,
have been evaluated~\cite{mt4,mt2,DGS}, and also the leading term of an
asymptotic expansion in the Higgs-boson mass has been obtained~\cite{mh2}.
The terms in the $\Mt$ expansion were found to be sizable, and the
next-to-leading term turned out to be about equally large as the
(formally) leading term~\cite{mt2}. Exact results have been derived for the
Higgs-mass dependence of the fermionic two-loop corrections to muon
decay~\cite{mhdepDeR}. In the present paper the Higgs-mass dependence
of $\De r$ (which is derived from muon decay~\cite{sirlin}), of the 
effective weak mixing angle at the Z-boson resonance, $\sweff$, and of the
leptonic width of the Z~boson, $\Ga_l$, is briefly discussed. The
results are compared with those obtained by
the expansion in the top-quark mass.
Furthermore the pure fermion-loop contributions, which constitute the
leading corrections in one-loop order, are analyzed in higher orders.
Exact results for these contributions to $\De r$ are derived up to
four-loop order. They are compared with the resummations of the leading
one-loop terms according to \citere{resum}.

\section{Higgs-mass dependence of $\De r$, $\sweff$ and $\Ga_l$ at two-loop
order}

In order to study the Higgs-mass dependence of the precision observables,
we consider subtracted quantities of the form 
\beq
a_{\mathrm{subtr}}(\MH) = a(\MH) - a(M^0_{\PH}), \quad \mbox{where }
a = \De r, \sweff, \Ga_l,
\eeq
and $\De r$ is defined by
$\MW^2 (1 - \MW^2/\MZ^2) = (\pi \al)/(\sqrt{2} \GF) (1 + \De r)$.
The subtracted quantities $a_{\mathrm{subtr}}(\MH)$ indicate the shift
in the precision observables caused by varying the Higgs-boson mass
between $M^0_{\PH}$ and $\MH$. In the analysis below we will consider
values of $\MH$ in the interval $65 \GeV \leq \MH \leq 1 \TeV$. Allowing
for such a wide range of Higgs-boson mass values may seem somewhat
over-conservative regarding the fact that the SM fits favor a light
Higgs-boson with a mass smaller than about 300~GeV~\cite{datamor98}.
However, the form of the fit curve and thus the value of the upper bound
is affected also by the SM predictions for larger $\MH$-values (and
also for $\MH$-values below the current experimental lower bound of
$\MH$). Thus
accurate theoretical predictions for the observables are needed in a
rather wide range of $\MH$-values in order to obtain a reliable fit
result.
 
Potentially large $\MH$-dependent contributions are the ones associated
with the top quark due to the large Yukawa coupling $\Pt \bar{\Pt} \PH$,
and the corrections proportional to $\De\al$. Further contributions are
the ones of the light fermions (except the corrections already contained
in $\De\al$), and purely bosonic contributions. The latter are expected
to be rather small~\cite{mhdepDeR}. We therefore have focussed on the 
Higgs-dependent fermionic contributions to the precision observables, 
for which we have obtained exact two-loop results~\cite{mhdepDeR,bsw}.

The methods used for these calculations have been outlined in 
\citere{sbaugw1}. The generation of the diagrams and counterterm 
contributions is performed with the help of computer-algebra
tools~\cite{CA}. For the renormalization the complete on-shell scheme
(with the conventions as in \citere{Dehab}) has been used, i.e.\
physical parameters are used throughout. The two-loop scalar integrals 
are evaluated numerically with one-dimensional integral
representations~\cite{intnum}.

\begin{figure}
\begin{center}
\epsfig{figure=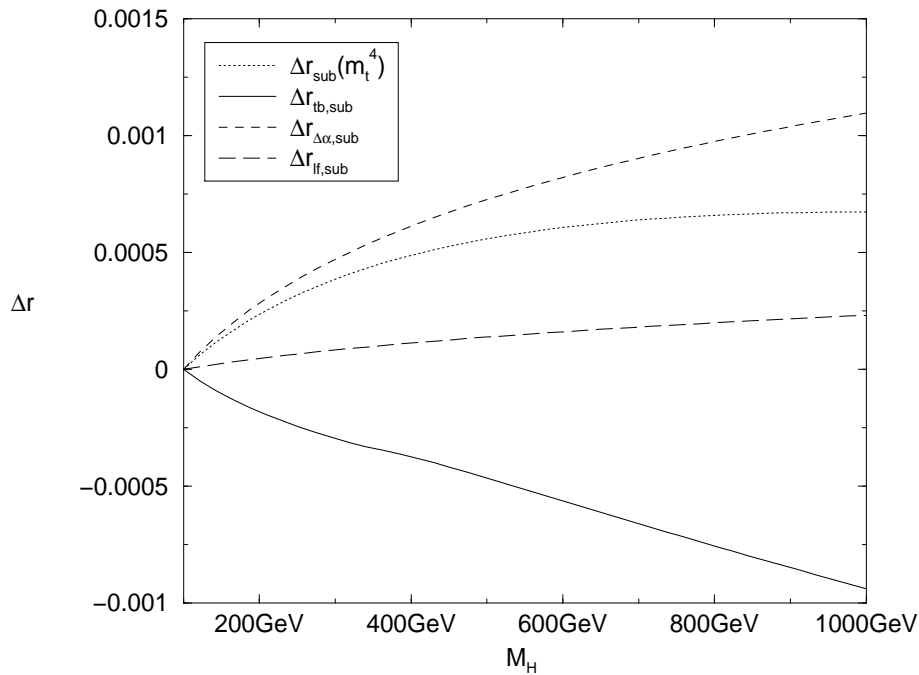, width=12cm, height=9cm}
\end{center}
\caption{\label{fig:drmh2l}
Higgs-mass dependent fermionic contributions to $\De r$ at two-loop
order. The different curves show the contribution from the diagrams
involving the top/bottom doublet ($\De r_{\mathrm{tbh}}$), the contribution
proportional to $\De\al$ ($\De r_{\De\al}$), the contribution of the
light fermions ($\De r_{\mathrm{lfh}}$), and the approximation of the
top/bottom correction by the leading term proportional to $\Mt^4$ 
($\De r(\Mt^4)$).
} 
\end{figure}

\reffi{fig:drmh2l} shows the Higgs-mass dependence of the two-loop
corrections to $\De r$ associated with the top/bottom doublet, with
$\De\al$, and with the light fermions. The dotted line furthermore 
indicates the Higgs-mass dependence of the leading $\Mt^4$-term~\cite{mt4}
in the top/bottom contribution. The top/bottom contribution gives rise
to a shift in the W-boson mass of $\De M^{\mathrm{top}}_{\PW,
\mathrm{subtr}, (2)}(\MH = 1000 \GeV) \sim 16$~MeV, which is about
$10\%$ of the one-loop contribution. Its Higgs-mass dependence turns
out to be very poorly approximated by the leading $\Mt^4$-term; the 
contribution of the latter even enters with a different sign. It can be seen
from \reffi{fig:drmh2l} that the two-loop contributions to a large
extent cancel each other. The contribution of the light fermions yields
a shift in $\MW$ of up to $\De M^{\mathrm{lf}}_{\PW,
\mathrm{subtr}, (2)}(\MH = 1000 \GeV) \lsim 4$~MeV. In total the
two-loop contributions lead to a slight increase in the sensitivity of
$\De r$ to the Higgs-boson mass compared to the one-loop case.
 

\begin{table}
$$
\begin{array}{|c||c|c||c|} \hline
\MH /\GeV &
\MWsub^{\mathrm{top}, \De\al} /\MeV &
\MWsub^{\mathrm{top}, \De\al, \mathrm{DGS}} /\MeV &
\De \MW /\MeV\\ \hline
65   & 0    & 0    & 0   \\ \hline
100  & - 22  & - 23  & 1 \\ \hline
300  & - 93  & - 98  & 5 \\ \hline
600  & - 145 & - 152 & 7 \\ \hline
1000 & - 183 & - 191 & 8 \\ \hline
\end{array}
$$
\caption{\label{tab:DeRcomp}
The Higgs-mass dependence of $\MW$ based on the exact result for the 
contribution of the top/bottom doublet (left column) and on the result 
of the expansion in $\Mt$~\cite{DGS} (right column).
}
\end{table}

In \refta{tab:DeRcomp} the Higgs-mass dependence of $\MW$ based on the
exact result for the top/bottom doublet is compared with the results of
the expansion in the top-quark mass up to ${\cal O}(\GF^2 \Mt^2 \MZ^2)$
given in \citere{DGS} (with the input parameters as in \citere{DGS}).
Besides the difference in the
two-loop results also higher-order effects due to differences in the
renormalization procedure and in the treatment of non-leading
higher-order corrections enter the comparison (in the comparison 
performed in \citere{mhdepDeR} the QCD corrections have been omitted). 
%
%
A discussion about different options for treating these
higher-order corrections will be given in \citere{bsw}.
Over the range of Higgs-mass values from 65~GeV to 1~TeV the difference
between the results amounts to about 8~MeV, which corresponds to $50\%$
of the two-loop top/bottom contribution. 
The difference in the Higgs-mass dependence of 8~MeV is slightly
larger than the theoretical uncertainty quoted in \citere{degrasstalk}
of 2~MeV from electroweak and 5~MeV from QCD corrections.
The contribution of the light fermions has not been included in the 
comparison in \refta{tab:DeRcomp}, since it is not contained in the
result of \citere{DGS}. As mentioned above, its Higgs-mass dependence
gives rise to a further shift of up to 4~MeV in the considered 
range of $\MH$-values.

We have performed an analogous analysis also for $\sweff$ and $\Ga_l$.
The detailed results will be presented in a forthcoming 
publication~\cite{bsw}. For the shift in
$\sweff$ induced by varying $\MH$ from 65~GeV to 1~TeV we find (leaving
out again the light-fermion contribution) $\sweffsub^{\mathrm{top},
\De\al}(\MH = 1 \TeV) = 13.5 \cdot 10^{-4}$, which differs from the
value given in \citere{DGS} by $1.0 \cdot 10^{-4}$. This difference 
amounts to about 80\% of the two-loop top/bottom contribution and is 
larger than the uncertainty quoted in \citere{degrasstalk} of 
$4 \cdot 10^{-5}$
(electroweak) and $3 \cdot 10^{-5}$ (QCD). It turns out that the
difference in the predictions for $\sweff(\MH)$ is mainly 
induced by the difference in the prediction for $\MW(\MH)$. We have 
calculated $\sweffsub^{\mathrm{top},\De\al}(\MH = 1 \TeV)$ using 
the value for $\MW(\MH = 1 \TeV)$ from \citere{DGS} instead of our
result given above, and found a value for
$\sweffsub^{\mathrm{top},\De\al}(\MH = 1 \TeV)$ that differs from the
one given in \citere{DGS} only by $2 \cdot 10^{-5}$.
The Higgs-mass dependence of the light-fermion contributions, which is
not included in \citere{DGS}, amounts to a shift in $\sweff$ of up to
$2 \cdot 10^{-5}$. For the Higgs-mass dependence of the leptonic width 
of the Z~boson, $\Ga_l$, we find good agreement with the result of the 
expansion in $\Mt$~\cite{mt2Gal}. The difference amounts to about 
$15 \%$ of the two-loop top/bottom contribution.

\section{Fermion-loop corrections}

The dominant contributions to the electroweak precision observables at
one-loop order are the fermion-loop corrections $\De \al$ and $\De\rho =
N_C \frac{\GF \Mt^2}{8 \pi^2 \sqrt{2}}$, which arise from the
renormalization of the electric charge and the weak mixing angle, 
respectively.
One of the main issues in improving the one-loop predictions has
therefore been to find a proper resummation of these leading one-loop
contributions. In the case of $\De r$, for instance, it could be 
shown~\cite{resum} that the resummation
\beq
1 + \De r \to \frac{1}{(1-\Da)(1+\frac{\cw^2}{\sw^2} \Drho) -
\Dri{rem}} 
\label{eq:resCHJSir}
\eeq
correctly takes into account the terms of the form
$(\De\al)^2$, $(\De\rho)^2$, $(\De\al \De\rho)$, 
$\De\al \De r_{\mathrm{rem}}$ 
at two-loop order. However, even with the knowledge of
the recently evaluated terms of 
${\cal O}(\GF^2 \Mt^2 \MZ^2)$~\cite{mt2,DGS} different treatments of 
contributions whose 
correct resummation is not known (i.e.\ non-leading two-loop terms and
the higher-order terms except the leading term in $\De\al$) may result
in predictions for $\MW$ that differ from each other by several MeV.

\begin{figure}[htb]
\begin{center}
\psfig{figure=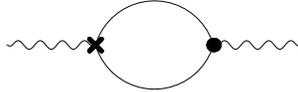, width=4cm,bbllx=74,bblly=520,bburx=222,bbury=580}
\end{center}
\caption{\label{fig:ctdiag}
A one-loop counterterm insertion into a one-loop diagram contributing
to the fermion-loop corrections at two-loop order.}
\end{figure}

We have investigated the pure fermion-loop contributions to
the precision observables, i.e.\ contributions containing $n$ fermion
loops at $n$-loop order (which in the case of quark loops are therefore
proportional to the $n$th power of the color factor $N_C$).%
\footnote{These contributions obviously constitute an UV-finite and
gauge-invariant subset.} 
The relevant diagrams in a given loop order are reducible diagrams of
vacuum-polarization type, which can easily be taken into account by
Dyson summation, but also graphs with counterterm insertions into
fermion-loop diagrams (see \reffi{fig:ctdiag}). We have derived
recursive relations which allow to express the $n$-loop
results in terms of one-loop one- and two-point functions~\cite{HKSW}.

Using the on-shell renormalization scheme we
have worked out explicit results for the fermion-loop contributions
to the precision observables up to four-loop order~\cite{HKSW}.
We have compared our result for the two-loop contribution to $\De r$
with the $\overline{\mbox{MS}}$ result derived in \citere{floopmsb}, 
and found very good agreement within less than 1~MeV for the $\MW$
prediction. Our exact results allow to investigate the validity of the
resummation \refeq{eq:resCHJSir} for the non-leading two-loop and the 
higher-order contributions. We found that already the first non-leading
two-loop term of ${\cal O}(\De\rho \log(\Mt^2/\MZ^2))$
is not correctly produced by the resummation \refeq{eq:resCHJSir}.
The terms of the form $\De\rho \De r_{\mathrm{top,rem}}$ 
and $\left(\De r_{\mathrm{top,rem}}\right)^2$
generated by \refeq{eq:resCHJSir} give rise to a deviation in $\MW$ of
about 7~MeV compared to the exact result for the top/bottom doublet.
It turns out, however, that large numerical cancellations occur between
the non-leading two-loop terms that are not correctly resummed in
\refeq{eq:resCHJSir}. While separately these terms
differ from the corresponding terms in the exact result by several MeV,
their sum approximates the full result remarkably well, up
to less than 1~MeV.

{}From our exact result for the three- and four-loop contributions we 
find that \refeq{eq:resCHJSir} in fact correctly produces the leading terms 
in $\De\al$, $\De\rho$ of the form $\left(\De\al\right)^a
\left(\De\rho\right)^b$, where $a + b = n$, and $n$ is the number of
loops. The three-loop term of the form $\left(\De\al\right)^2 \De
r^{\al}_{\mathrm{ferm, rem}}$ is also correctly generated, while 
the three-loop term $\De\al \De r^{\al^2}_{\mathrm{ferm, rem}}$
is not correctly produced. In the full fermion-loop contribution at
three-loop order again accidental numerical cancellations occur, which
give rise to the fact that the total contribution is much smaller than
the individual terms and affects the prediction for $\MW$ by less than
1~MeV. 

In total we find that due to accidental numerical cancellations
the resummation of the fermion-loop contributions according to 
\refeq{eq:resCHJSir} yields a very good numerical approximation
of the complete fermion-loop result. The exact result is approximated
within about 2~MeV. The results for the pure fermion-loop contributions 
to $\sweff$ and $\Ga_l$ will be discussed in \citere{HKSW}.\\

%
%
%

The author thanks T.~Riemann and the other organizers of the Zeuthen
Workshop for their kind hospitality and the pleasant atmosphere during
the workshop. I am grateful to my collaborators
S.~Bauberger, W.~Hollik, B.~Krause and A.~Stremplat,
with whom the results presented here have been worked out, and to 
G.~Degrassi, P.~Gambino and A.~Sirlin for useful discussions.

\end{document}